

\input harvmac

\Title{BROWN-HET-888}{A Proposal on the Topological Sector of 2d String}
\centerline{Miao Li\foot{E-mail: li@het.brown.edu}}
\bigskip
\centerline{Department of Physics}
\centerline{Brown University}
\centerline{Providence, RI 02912}
\bigskip
The field content of the two dimensional string theory consists of the
dynamical tachyon field and some nonpropagating fields which consist
in the topological sector of this theory. We propose in this paper to
study this topological sector as a spacetime gauge theory with a simple
centrally extended $w_\infty$ algebra. This $w_\infty$ algebra appears in
both the world sheet BRST analysis and the matrix model approach. Since
the two dimensional centrally extended Poincar\'e algebra is naturally
embedded in the centrally extended $w_\infty$ algebra, the low energy
action for the metric and dilaton appears naturally when the model is
truncated at this level. We give a plausible explanation of emergence
of discrete states in this formulation. This theory is again the effective
theory at zero slope limit. To include higher order $\alpha'$ corrections,
we speculate that the whole theory is a gauge theory of a deformed
$w_\infty$ algebra, and the deformation parameter is just $\alpha'$.

\Date{1/93}

\newsec{Introduction}

Despite much progress has been made in the last few years in the study
of the two dimensional string theory \ref\kle{For a review see I.R.
Klebanov, ``Strings in two dimensions'', Lectures at ICTP Spring School
on String Theory and Quantum Gravity, Trieste, April 1991.}, many
issues remain open. The matrix model approach provides us a very powerful
tool in calculating scattering amplitudes of the tachyon \ref\ampl{K.
Demeterfi, A. Jevicki and J.P. Rodrigues, Nucl. Phys. B362 (1991) 173;
J. Polchinski, Nucl. Phys. B362 (1991) 125; G. Moore, Nucl. Phys.
B368 (1992) 557; G. Mandal, A.M. Sengupta and S.R. Wadia, Mod. Phys.
Lett. A6 (1991) 1465.}, sometimes even nonperturbatively \ref\nonp{
G. Moore, R. Plesser and S. Ramgoolam, Nucl. Phys. B377 (1992) 143.}.
Just how the perturbatively unitary S-matrix \nonp\ can be understood in
a conventional spacetime picture is evading our grip so far. There
are many reasons for one to go beyond what we have achieved. Although
discrete states discovered in both the conformal approach \ref\dis{
A.M. Polyakov, Mod. Phys. Lett. A6 (1991) 635; B. Lian and G.
Zuckerman, Yale preprint YCTP-P18-91; P. Bouwknegt, J. McCarthy and
K. Pilch, CERN preprint CERN-TH.6162/91.} and the matrix
model approach \ref\mat{D.J. Gross, I.R. Klebanov and M.J. Newman,
Nucl. Phys. B350 (1991) 621.} are nonpropagating, they play an important
role in providing more backgrounds such as a black hole background and also
in the tachyon S-matrix. The latter is unitary with a tachyon background,
thus discrete states should neither appear in the intermediate channels
and nor serve as external scattering states, at least in this
background. Nevertheless a recent calculation \ref\li{M. Li, Brown
preprint BROWN-HET-887, Dec. 1992.}
showed that the conventional spacetime effective action for the tachyon
is nonlocal, this nonlocality may be understood by integrating out
nonpropagating modes. Precisely in this place one need to understand how
the tachyon is coupled to other topological modes and how these modes are
integrated out to result in the effective action for tachyon.

To study the topological sector in this theory and its coupling to the tachyon,
a full formulation such as a string field theory is needed. Unfortunately
although much progress in that field has been made recently
\ref\barton{B. Zwiebach, IAS preprint IASSNS-HEP-92/41.}, we are still
lacking a tractable string field theory. Our only guide here is the fact
that the theory
for the topological sector must be topological, so its form can not be
random. A natural formulation of a topological theory is a gauge theory of
some group or algebra. Within such a formulation,
the $w_\infty$ symmetries found in the 2d string in different
ways \ref\symm{J. Avan and A. Jevicki, Phys. Lett. B266 (1991) 35;
D. Minic, J. Polchinski and Z. Yang, Nucl. Phys. B369 (1992) 324; S. Das,
A. Dhar, G. Mandal and S. Wadia, Mod. Phys. Lett. A7 (1992) 71.}
\ref\witt{E. Witten, Nucl. Phys. B373 (1992) 187; I.R. Klebanov and A.M.
Polyakov, Mod. Phys. Lett. A6 (1991) 3273; E. Witten and B. Zwiebach,
Nucl. Phys. B377 (1992) 55.} should be the residual symmetries of gauge
symmetries after a certain gauge fixing.

A somewhat straightforward approach is
to gauge the $w_\infty$ algebra in spacetime, this is what to be done in
this paper. In the matrix model approach, $w_\infty$ algebra arises due to
the free fermion nature of the theory, and it also appears as an infinite
algebra commuting with the collective field hamiltonian. This very same
algebra arises on the world sheet as the algebra of spin one vertex operators,
having direct relation to discrete states in the BRST cohomology. Spacetime
momenta are discrete for these currents. In both approaches,  spacetime
origin of the $w_\infty$ algebra is missing. To motivate our formulation,
we should look for such a spacetime interpretation. There is a ready
interpretation, which is just to interpret the $w_\infty$ as the
area-preserving diffeomorphism algebra of the plane, the
spacetime. 2d Poincar\'e algebra (Euclidean algebra in case of Euclidean
spacetime) is a subalgebra of this $w_\infty$ algebra. This subalgebra should
play a peculiar role in the story. Indeed H. Verlinde
in \ref\ver{H. Verlinde, in Sixth Marcel Grossmann Meeting on General
Relativity, M. Sato ed. (World Scientific, Singapore, 1992).} proposed
to study the low energy action of the dilaton and the metric as a gauge
theory of the 2d Poincar\'e algebra. The topological nature of the theory
is obvious in this formulation. Cangemi and Jackiw later reformulated this
theory as a gauge theory of the centrally extended Poincar\'e algebra
\ref\jack{D. Cangemi and R. Jackiw, MIT preprint CTP\#2147, 1992.}. The
cosmological constant term has a natural origin in this formulation.
This gauge formulation is another motivation for our search for a gauge
theory of the whole topological sector, in which the dilaton and metric
are just the lowest modes. It turns out that the central extension of
Poincar\'e
algebra can be naturally embedded into the corresponding central extension
of $w_\infty$ algebra, and the center element has a natural position in
the latter algebra.

In the next section we shall briefly review the gauge formulation of the
metric and dilaton system, and point out an important point which becomes
crucial in the $w_\infty$ extension, in the end of that section. We then in
section 3 introduce the
$w_\infty$ algebra as the area-preserving diffeomorphism algebra of the
plane, and construct the gauge theory. In section 4, classical solutions
are discussed and discrete states are explained in our theory. We
present our conclusion and outlook in the final section.

\newsec{Dilaton plus the metric as a gauge theory}

We review in this section the observation made by H. Verlinde on the two
dimensional dilaton and metric system \ver, which is polished later by Cangemi
and Jackiw \jack. We will work in the Euclidean spacetime, rotating back to
the Minkowski spacetime is straightforward. The (zero slope) low energy
action for the dilaton and the metric in the 2d string theory is
\eqn\dm{S=\int e^{-2\Phi}\sqrt{g}(R-\partial \Phi\overline{\partial}\Phi
+\lambda),}
where we have used the complex coordinates system. $\lambda$ is the spacetime
cosmological constant, and is fixed in the string theory by requiring the
critical central charge on the world sheet. We shall leave the cosmological
constant arbitrary. Our convention for the curvature is, in the conformal
gauge,
$$\sqrt{g}R=\partial\overline{\partial}\hbox{ln}g_{z\bar{z}}.$$
This model allows for a family of black hole solutions parametrized by the
mass of the black hole \ref\black{S. Elitzur, A. Forge and E. Rabinovici,
Nucl. Phys. B359 (1991) 581; G. Mandal, A. Sengupta and S.R. Wadia, Mod.
Phys. Lett. A6 (1991) 1685.} \ref\witten{E. Witten, Phys. Rev. D44 (1991)
314.}. Any theory which admits solutions parametrized by a finite number
of moduli must be topological. H. Verlinde observed that in the first
order formalism, model \dm\ can be interpreted as a gauge theory with
the two dimensional Poincar\'e group as the gauge group and is obviously
topological. The cosmological constant term is gauge invariant on-shell,
namely when the field strength associated with the zwei-bein field vanishes.
Cangemi and Jackiw promoted this theory to be gauge invariant off-shell
by introducing a center element in the gauge algebra, thus the dilaton-metric
system is a gauge theory of the centrally extended Poincar\'e group. In
the Euclidean space time, the relevant generators are $(J, L_{-1},
\bar{L}_{-1},I)$. Here $I$ is the center element, $J$ is the generator of
rotations, and $L_{-1}$, $\bar{L}_{-1}$ are translation generators, in
the complex coordinates. Nonvanishing commutators are
\eqn\comm{\eqalign{&[J, L_{-1}]=iL_{-1},\qquad [J, \bar{L}_{-1}]=-i\bar{L}
_{-1}, \cr
&[\bar{L}_{-1}, L_{-1}]=iI.}}
In terms of the generators of the Virasoro algebra, $J$ can be written as
$i(L_0-\bar{L}_0)$. This fact is irrelevant for us, since we will not embed
the central extension of Poincar\'e algebra into a sub-algebra of the Virasoro
algebra. Next we introduce the gauge field
$$A=\omega J+eL_{-1}+\bar{e}\bar{L}_{-1}+aI$$
with the reality condition $e^*=\bar{e}$, $*$ is complex conjugation. The
field strength is then
$$F=dA+A^2=d\omega J+DeL_{-1}+D\bar{e}\bar{L}_{-1}+(da-ie\wedge \bar{e})I,$$
where covariant derivatives are defined as $De=de+i\omega\wedge e$ and
$D\bar{e}=d\bar{e}-i\omega\wedge\bar{e}$.
Introduce lagrangian multipliers $(X^0,\bar{X},X,\lambda)$ which transform
coadjointly under the centrally extended algebra. The action
\eqn\fir{\int \left(X^0d\omega+\bar{X}De+XD\bar{e}+\lambda(da-ie\wedge \bar{e})
\right)}
is gauge invariant. Equations of motion for the gauge field derived from \fir\
are the condition for flat connections. Equations of motion for the Lagrangian
multipliers are
\eqn\lag{\eqalign{&DX+i\lambda e=0,\qquad D\bar{X}-i\lambda \bar{e}=0,\cr
&dX^0+i\bar{X}e-iX\bar{e}=0, \qquad d\lambda=0.}}
The last equation implies that $\lambda$ is a constant. Substituting these
equations into \fir\ and using the following definition
$$X^0=e^{-2\Phi},\qquad g=e^{2\Phi}e\otimes\bar{e}$$
we recover the action \dm\ for the dilaton and the metric. Notice that the
rescaled metric $\hbox{exp}(-2\Phi)g$ is always flat by equation of motion.
Equations in \lag\ imply that  the combination $\lambda X^0-X\bar{X}$ is
a constant. Let it be $M$, then the metric and dilaton are
\eqn\sl{g={1\over\lambda(M+X\bar{X})}DXD\bar{X},\qquad e^{-2\Phi}={1\over
\lambda}(M+X\bar{X}).}
Suppose $g$ is nondegenerate, then $X$ and $\bar{X}$ can be chosen as a
complex coordinates system. We shall explain why this is the case shortly.
When this is done, the solution is precisely the Euclidean black hole
solution with mass $M$ found in \black\ \witten.

There is a crucial point ignored in the previous discussion in \ver. The
equation of motion for gauge field is the condition of flat connection. Since
the topology of the plane is trivial, one expects that any flat connection
can be gauge transformed into zero. If so, the metric in \sl\ would be
degenerate and we lose the black hole solutions. To solve the puzzle, we
should note that all solutions to the equation of motion for the gauge
field are gauge equivalent to one and another. The null solution and constant
solutions are particular solutions. For the latter case, we mean that
components in $e=e_zdz+e_{\bar{z}}d\bar{z}$ are constant. Take, say
$\omega=e_{\bar{z}}=0$ and $e_z$ nonvanishing, we see that following from
\lag\ $X$ is proportional to $z$. This gives the black hole solution.
When however all components of the gauge field are zero, solutions of
the lagrangian multipliers are constant solutions. Since there are four
such fields, it appears that there are solutions of four parameters.
This is not the case. When we fix the gauge with vanishing gauge field,
apparently there are residual gauge transformations. These gauge
transformations are just constant transformations, and there are four
parameters here too. It is easy to see that both $\lambda X^0-X\bar{X}$
and $\lambda$ are invariant under a constant gauge transformation. The first
quantity is just the parameter $M$ and the second, unfortunately, is fixed
by hand in our formulation.

The above argument still does not give us a satisfactory answer to why
we should favor a constant solution. The best way to settle this
problem is to couple the gauge system to a matter field, say the tachyon
in the context of 2d string theory. We leave this problem to a future
work.

To reproduce discrete ``states'' in our extended theory, we will encounter
exactly the same problem. Again the geometric origin is associated with
coupling the gauge theory of $w_\infty$ to the tachyon. Put in another
way, it should follow from a complete string field theory.

\newsec{Gauge theory of the centrally extended $w_\infty$}

We are going to show that the centrally extended Poincar\'e algebra can be
naturally embedded into a centrally extended $w_\infty$ algebra. For simplicity
and reason becoming clear later, we simply denote such a centrally extended
algebra also by $w_\infty$. In fact generators of the central extended
Poincar\'e algebra are the first few generators of $w_\infty$. We shall
discuss a gauge theory of this infinite dimensional algebra, along the
line in the previous section. This theory is again topological. When the
theory is truncated to the centrally extended Poincar\'e algebra which is
generated by the first few generators of $w_\infty$, the topological theory
of the dilaton and the metric is obtained. This is the first hint that this
theory might be the right theory
for the topological sector of the 2d string. We then proceed to show that
discrete states found previously in the conformal approach also naturally
emerge as classical solutions in a certain gauge. This is the second
evidence for taking this theory as a candidate theory for the topological
sector. We begin with discussing the $w_\infty$ algebra as the area-preserving
diffeomorphism algebra of the complex plane, i.e. the spacetime in the context
of the 2d string theory, in the next subsection.

\subsec{Area-preserving diffeomorphism algebra of the plane}

Our motivation for discussing such an algebra is the following. First, the
2d Poincar\'e algebra must be a sub-algebra of the area-preserving
diffeomorphism algebra, as rotations and translations obviously preserve
the area element. Second, precisely such algebra appears in both world
sheet analysis and the matrix model approach, where the $w_\infty$ symmetries
seem to be the residual symmetries after certain gauge fixing in the full
string theory. We are motivated to uncover the unknown, larger gauge
symmetries and identify the known $w_\infty$ symmetries as the residual
symmetries.

We continue to use complex coordinates. The area element is ${1\over i}dz
\wedge d\bar{z}$. Any diffeomorphism preserving this area element can
be generated by $\delta z=i\overline{\partial}f$, $\delta\bar{z}=-i\partial
f$, $f$ is a regular real function. Any such function can be expanded into
series
$$f=\sum_{n,m\ge 0}f_{m,n}z^m\bar{z}^n$$
with reality condition $f^*_{m,n}=f_{n,m}$, in other words, $f=(f_{m,n})$
is a hermitian matrix. Note that a constant function results in a trivial
diffeomorphism. Nevertheless precisely this generator can be interpreted
as a center element, as explained later. For a pair of functions $f$ and
$g$, we define the Poisson bracket associated to our area element:
$$\{f, g\}=i(\partial f\overline{\partial}g-\overline{\partial}f\partial g)$$
For the basis $V_{m,n}=z^m\bar{z}^n$, we have a closed algebra
\eqn\basis{[V_{m_1, n_1}, V_{m_2,n_2}]=i(m_1n_2-m_2n_1)V_{m_1+m_2-1,
n_1+n_2-1},}
where we have replaced the Poisson bracket by commutator, for later
convenience. This algebra is just the $w_\infty$ algebra, as becomes
familiar in a new basis $Q_{s,l}=V_{(s+l)/2, (s-l)/2}=r^s\hbox{exp}(il\theta)$.
In the second expression we used the polar coordinates $z=r\hbox{exp}
(i\theta)$. Note that for nontrivial generators $s\ge 1$, the range of
integer $l$ for a given $s$ is $-s, -s+2,\dots, s$. Commutators in this basis
are
\eqn\winf{[Q_{s_1,l_1}, Q_{s_2,l_2}]={i\over 2}(l_1s_2-l_2s_1)Q_{s_1+s_2-2,
l_1+l_2}.}
This is precisely the $w_\infty$ algebra found in the 2d string theory. It is
easy to identify the translation generators and the rotation generator,
from our discussion. Indeed $L_{-1}=Q_{1,-1}$, $\bar{L}_{-1}=Q_{1,1}$ and
$J=Q_{2,0}$. These are first few generators in $w_\infty$. The Euclidean
subalgebra is
\eqn\sub{\eqalign{&[Q_{2,0},Q_{1,-1}]=iQ_{1,-1},\cr
&[Q_{2,0},Q_{1,1}]=-iQ_{1,1},}}
to be compared with commutators in \comm. Remarkably enough, we can extend
the algebra \winf\ to include the element $Q_{0,0}$. This element might
be explained as the trivial generator corresponding to the constant function
$f$. Commutation relation in \winf\ extended to include $Q_{0,0}$ is
consistent, and we have
$$[Q_{1,1},Q_{1,-1}]=iQ_{0,0}.$$
Compared to \comm\ we soon find the identification $Q_{0,0}=I$. Note that,
according to \winf\ this generator commutes with all other generators. We
thus find that the center element in the centrally extended Euclidean
(Poincar\'e) algebra can be very naturally included in the $w_\infty$ algebra.
We shall denote this extended algebra again by $w_\infty$.

Finally, we note that the $w_\infty$ algebra is different from the algebra
of area-preserving diffeomorphism on a cylinder or a punctured plane
\ref\shen{For a review see
X. Shen, Int. J. Mod. Phys. A7 (1992) 6953.}. A copy of Virasoro algebra
can be embedded in the latter, but not in the $w_\infty$ algebra under
discussion.

\subsec{The gauge theory}

It is straightforward to generalize the gauge theory discussed in section 2
to the whole $w_\infty$ algebra introduced in the previous subsection. The
gauge field is
\eqn\field{A=\sum_{s,l}A^{s,l}Q_{s,l}.}
Reality condition $(A^{s,l})^*=A^{s,-l}$ is imposed. By embedding of the
centrally extended Poincar\'e algebra, we identify the first few components
\eqn\iden{\eqalign{&A^{0,0}=a,\qquad A^{2,0}=\omega,\cr
&A^{1,-1}=e,\qquad A^{1,1}=\bar{e}.}}
The field strength, calculated from the structure of the algebra, is
\eqn\stren{\eqalign{&F=dA+A^2=\sum_{s,l}F^{s,l}Q_{s,l},\cr
&F^{s,l}=dA^{s,l}+\sum_{s',l'}{i\over 4}(l's-ls'+2l')A^{s',l'}\wedge
A^{s-s'+2,l-l'}.}}
In particular, the lowest components are identified with those of the central
extension of Euclidean algebra
\eqn\low{\eqalign{&F^{0,0}=da-ie\wedge\bar{e},\qquad
F^{1,-1}=de+i\omega\wedge e +2i\bar{e}\wedge A^{2,-2}\cr
&F^{1,1}=d\bar{e}-i\omega\wedge\bar{e}-2ie\wedge A^{2,2}\cr
&F^{2,0}=d\omega-ie\wedge A^{3,1}+i\bar{e}\wedge A^{3,-1}+4iA^{2,2}\wedge
A^{2,-2}.}}
Two notable new features appear in these components. First, the spin
connection is not simply related to zwei-bein fields when condition $F^{1,-1}=
F^{1,1}=0$ is imposed, because of new components in the gauge field. This
implies that the spin connection has torsion. Second,
again due to new components, the spin connection is no longer flat when
the gauge field is a flat connection. One is attempting to interpret this
as back-reaction of high rank fields to the spin connection and zwei-bein
fields.

Introducing lagrangian multipliers $(X_{s,l})$ which transform coadjointly
under gauge transformations and imposing the reality condition $(X^{s,l})^*
=X^{s,-l}$, we generalize the action presented in section 2:
\eqn\new{S=\int\sum_{s,l}X_{s,l}F^{s,l}.}
This is a topological theory too. The equation of motion for the gauge
field is simply $F^{s,l}=0$. For a component $F^{s,l}$, there are only
finitely many components of the gauge field involved. The identification
of lowest $X_{s,l}$ with those discussed in section 2 is
$$\eqalign{&X_{0,0}=\lambda,\qquad X_{1,-1}=\bar{X},\cr
&X_{1,1}=X,\qquad X_{2,0}=X^0.}$$
The equation of motion for lagrangian multipliers is
\eqn\mul{dX_{s,l}+\sum_{s',l'}(l's-ls'+2l)X_{s',l'}A^{s'+2-s,l'-l}=0.}
Note that for a given $(s,l)$, there are infinitely many components of
$X$ and $A$ involved in \mul. We can not derive simple relations
among $X^{s,l}$
as integrations of the equation of motion. In addition, it is not possible to
use the equation of motion to eliminate the lagrangian multipliers in the
action \new.

\newsec{Classical solutions}

The equation of motion for the gauge field is the condition of flat
connection. Just as in the theory discussed in section 2, any such
connection is gauge equivalent to the vanishing one, as we choose
spacetime as the 2-plane. Does this imply that the content of our theory is
trivial? The answer is no. Suppose we are allowed to gauge transform
the gauge field to zero, the residual gauge transformations are then
constant gauge transformations. The equation of motion for $X_{s,l}$ in
this gauge is just $dX_{s,l}=0$. So solutions of $X_{s,l}$ are constant
solutions, and the space of solutions is infinite dimensional. Since the
residual gauge transformations are constant, the space of gauge inequivalent
solutions is the space of coadjoint orbits of the 2d area-preserving
diffeomorphism group in the infinite dimensional space of constant
solutions of $(X_{s,l})$. Note that, the central element has no
impact in a gauge transformation.

Under gauge transformation with gauge parameter $\epsilon=\sum\epsilon^{s,l}
Q_{s,l}$, the gauge field transforms according to
$$\delta A^{s,l}=d\epsilon^{s,l}+\sum_{s',l'}{i\over 2}(l's-ls'+2l')A^{s',l'}
\epsilon^{s+2-s',l-l'},$$
and the lagrangian multipliers transform as
\eqn\trans{\delta X_{s,l}=-\sum_{s',l'}{i\over 2}(l's-ls'+2l)X_{s',l'}\epsilon
^{s'+2-s,l'-l}.}
The gauge transform of $X_{s,l}$ is essentially different from that of
the gauge field in that there are infinitely many terms in \trans.

We claim that the space of gauge inequivalent solutions is also infinite
dimensional. To see this, consider a solution in which all components
of $X$ are zero except for $X_{0,0}$, $X_{s,l}$ and $X_{s,-l}=(X_{s,l})^*$.
$X_{0,0}$ is the cosmological constant $\lambda$ and we have to fix it in
the string theory. It is invariant under gauge transformation \trans\ for any
configuration of $X$. Now we show that the above particular configuration
can not be transformed into zero. We have $\delta X_{s,l}=-il\lambda
\epsilon^{2-s,-l}-ilX_{s,l}\epsilon^{2,0}$. If $s$ is greater than 1, the
first term is zero. The second term does not change the absolute value of
$X_{s,l}$, since $\epsilon^{2,0}$ is real so that the second term is a pure
phase shift. It is easy to see that any such two distinct configurations
with different pairs $(s,l)$ and $(s',l')$ are not gauge equivalent.
This shows that the space of gauge inequivalent solutions is infinite
dimensional.

Recall that in the conformal analysis, there are infinitely many new physical
``states'' found in \witt, and each state is assigned with a pair of $(s,l)$
\foot{The new states found in the third paper in \witt\ are argued in
\ref\muk{S. Mahapatra, S. Mukherji and A.M. Sengupta, Mod. Phys. Lett.
A7 (1992) 3119.} to be physically trivial, as can be eliminated by gauge
transformations in string field theory.}. Each state of this kind has
pure imaginary energy in Minkowski space in which the Liouville dimension
is spatial, therefore it is nonpropagating. However, these states can be used
to deform the world sheet conformal field theory, thereby provide new
backgrounds. Given that the space of solutions in our theory is infinite
dimensional, and the solution $(\lambda, X_{s,l}, X_{s,-l})$ is a nontrivial
solution, we conjecture that this space is actually the moduli space of all
possible physically inequivalent backgrounds in the 2d string, when the
tachyon is absent. In particular,
the solution $(\lambda, X_{s,l}, X_{s,-l})$ may be identified with the moduli
corresponding to deformation generated by corresponding pair of discrete
states. This is encouraged by the simplest case of the black hole solution,
which is gauge equivalent in the present theory to $(\lambda, X_{2,0}=X^0)$.

As we discussed in section 2, one should not be satisfied with the degenerate
solution, although in the pure topological theory it is equivalent to the
nondegenerate solution, namely the black hole solution. All solutions we
discussed above are degenerate, as the gauge field is zero.
Nevertheless it is
tempting to transform a degenerate solution into a nondegenerate one.
As suggested by the constant solution considered in section 2, a particularly
attractive nondegenerate solution is the following. We still assume
the zwei-bein field $e=e_zdz$, and $e_z$ is a nonvanishing constant.
Thus $e_z$ is scalar
under rotations, this corresponds to asking the gauge field $A$ be scalar
under rotations, since the generator $L_{-1}$ transforms in the opposite way
as $dz$ does. Now consider component $A^{s,-l}$ with $l>0$. We assume
the gauge $A^{s,-l}=A^{s,-l}_zdz$. To require also that $A$ is scalar under
rotations, then $A^{s,-l}_z$ takes a form $f(r)\hbox{exp}(i(l-1)\theta)$ in
the polar coordinates. This form is very suggestive, if we take a special
function $f(r)=r^s$. For large $r$, the dilaton field $X^0=\hbox{exp}(-2\Phi)
\rightarrow r^2$. In the linear dilaton region, we identify
$r=\hbox{exp}(-\phi)$,
where $\phi$ is the Liouville dimension. Using this identification we find
$A^{s,-l}_z=\hbox{exp}(-s\phi+i(l-1)\theta)$. This form is the same as the
momentum factor in the vertex operator of a discrete state. Note that, the
oscillator part of the vertex operator depends on coordinates of the world
sheet, therefore has nothing to do with a spacetime field.

We point out that when $A^{s,l}$ is nonzero, some higher rank components
must be nonzero too. Suppose $A^{s,l}$ is the first nonvanishing component
of the gauge field, except for $e$ and $\bar{e}$, then the equation of
motion for $A^{s,l}$ is
$$dA^{s,l}-{i\over 2}(s+l+2)e\wedge A^{s+1,l+1}+{i\over 2}(s-l+2)\bar{e}
\wedge A^{s+1,l-1}=0.$$
If $A^{s,l}$ takes the above suggested form, we find that the last two terms
in the above equation can not be zero simultaneously. This can be simply
understood. The vertex operator corresponding to $A^{s,l}$ is not exactly
marginal on the world sheet, so must be supplemented with other terms on the
world sheet.

As suggestive as it appears, the above form of the gauge field $A^{s,-l}$,
if has anything to do with the known discrete state, must be justified in a
more complete theory. As $e$ must be nondegenerate if the tachyon is coupled
to the gauge system, the form of a higher rank component may also be determined
in a similar way. Finally, we note that although we argued that the
space of physical solutions is infinite dimensional, we have not probed
its structure extensively. As the space of coadjoint orbits of the 2d
area-preserving diffeomorphism group, it can be studied in a mathematically
rigorous way.

\newsec{Conclusion and outlook}

We have proposed in this paper a simple topological theory as the theory of
the topological sector in the 2d string theory. This is a straightforward
and natural generalization of the gauge theory for the system of the dilaton
and metric. Our analysis in the last section is especially encouraging,
since discrete states in principle can be understood in this formulation.
There are many things remaining to do. The most pressing one is to couple
this system to the tachyon and study the whole system both at classical
and quantum levels.

Another issue we leave open in this paper is that of higher order corrections
in $\alpha'$. This parameter finds no place in the present theory. It is
certainly important to incorporate these corrections into the theory,
in order to understand some genuine stringy phenomena. For example when the
curvature becomes large, these corrections become significant. It is
especially interesting to understand the exact black hole solution
suggested in \ref\dvv{R. Dijkgraaf, H. Verlinde and E. Verlinde, Nucl.
Phys. B371 (1992) 269.}. We believe that the topological sector remains
topological, when higher order $\alpha'$ corrections are included. The
theory may also be formulated as a gauge theory. There is a ready choice,
namely to use some deformation of $w_\infty$ algebra as the gauge algebra
in which $\alpha'$ is the deformation parameter. This possibility is currently
under investigation.

\vskip1cm
\noindent{Acknowledgements}

I am grateful to A. Jevicki for inspiration. I am also grateful to Y.S. Wu
for bringing \jack\ to my attention. This work was supported by DOE contract
DE-FG02-91ER40688-Task A.

\listrefs\end